\documentstyle[twocolumn, epsfig, epstopdf]{article}

\makeatletter
\def\@normalsize{\@setsize\normalsize{10pt}\xpt\@xpt
\abovedisplayskip 10pt plus2pt minus5pt\belowdisplayskip
\abovedisplayskip \abovedisplayshortskip \z@
plus3pt\belowdisplayshortskip 6pt plus3pt
minus3pt\let\@listi\@listI}
\def\subsize{\@setsize\subsize{12pt}\xipt\@xipt}
\def\section{\@startsection {section}{1}{\z@}{1.0ex plus
1ex minus .2ex}{.2ex plus .2ex}{\large\bf}}
\def\subsection{\@startsection {subsection}{2}{\z@}{.2ex
plus 1ex} {.2ex plus .2ex}{\subsize\bf}} \makeatother

\begin{document}

\title{A XML Based Datagrid Description Language}

\author{Rene Felder and Erich Schikuta\\
Institut f\"ur Informatik und Wirtschaftsinformatik, University of Vienna\\
Rathausstr. 19/9, A-1010 Vienna, Austria\\
erich.schikuta@univie.ac.at}

\date{}

\maketitle

\begin{abstract}
\small{We present xDGDL, an approach towards a concise but
comprehensive Datagrid description language. Our framework is
based on the portable XML language and allows to store syntactical
and semantical information together with arbitrary files. This
information can be used to administer, locate, search and process
the stored data on the Grid.
As an application of the xDGDL
approach we present ViPFS, a novel distributed file system
targeting the Grid.}

\end{abstract}

\subsection*{Keywords:}
\small{Datagrid, XML, meta information, distributed file systems,
parallel and distributed I/O}

\section{Introduction}\label{Intro}

Today new and stimulating data-intensive problems in biology,
physics, astronomy, space exploration, human genom research arise,
which bring new  high-performance applications with the need  to
store, administer and search intelligently gigantic data set spread
over globally distributed storage resources \cite{segal00}.

We face a similar situation as in the well-known area of database
systems \cite{silberschatz96}, where data represents a model of the
reality. Information must be searched, analyzed, administered easily
and must be at hand efficiently for arbitrary applications.
Consequentially data has to be attributed with meta information
describing the specific semantics of the information in a
standardized and processable way. This meta information allows
applications to search the stored information intelligently.

However meta information in the context of Grid computing has to
describe not only the logical part of the data (semantical
information) but also specific structural information on the
physical distribution of the data (syntactical information). Thus
we propose an approach for an XML based language to act as a
notational tool to describe all this information for data stored,
administered, searched and processed on the Grid. Any information
stored on the Grid (from a conventional text file to a structured
database relation) is attributed with a sematic description
expressed by the XML notation. In the most simple case the XML
description is stored together with the file.

Only a few similar approaches exist, but these are in a early state
(e.g. \cite{feitelson01}) or target mostly very specific application
domains (e.g. \cite{bhatti00} \cite{hdf}).

The layout of the paper is as follows. In the next section we present
xDGDL, the XML-based Data Grid Description Language, and give several
examples for the usage of the language. Then we introduce shortly the
Meta-ViPIOS system \cite{fuerle00}, which is a client server based
I/O system supporting distributed applications on the Grid. Finally
we present an prove-of-concept implementation of the xDGDL language
within the ViPFS, the distributed file system component of the ViPIOS
system.

\section{xDGDL - the XML Data Grid Description Language}\label{xml_descriptor}

We propose the XML Data Grid Description Language (xDGDL) which aims
to provide a convenient XML framework for the specification of meta
information of data stored on the Grid. xDGDL is a derivative of
PARSTORAGE \cite{parstorage}, which was specifically designed as meta
language for parallel IO data.

The xDGDL descriptor consists of a logical and a physical view to the
file. The logical view describes the semantical information and the
physical view the syntactical information (the physical layout) of
the file.

Focusing the Grid we have to specify a very general Grid architecture
hosting our framework. From our point of view the Grid consists of an
arbitrary number of {\em collaborations}, which are defined by an
organizational domain \cite{foster-physio}, interconnected by WAN
technology. In practice such a collaboration will be usually (but
must not be) a coherent IT infrastructure represented by a cluster
like system, which consists of a number of execution nodes. These
{\em nodes} are {\em processing nodes} and/or {\em data (server)
nodes}. The latter type provides data storage resources by a number
of storage {\em devices} (e.g. disks, tapes, etc.). It is to note
that a single data node can host an arbitrary number of devices.

\subsection{The goals of xDGDL}

The basic idea of the XML based approach is quite simple: Together
with any "chunk" of data a xDGDL description of the meta information
of the data is stored, in other words, any arbitrary number of bytes
stored within our framework is attributed with its describing
information, delivering the following properties:

\subsubsection{Semantics of data}
Applications write results to files. There are lots of applications,
there are lots of formats, there are lots of files. But what can be
found in these files? Generally applications do not write simple
bytes into a file. They write integers, real numbers, characters,
records of arbitrary types etc. So the contents of a file is not just
a sequence of bytes, but it is a sequence of typed elements. Without
the knowledge of the semantics of the applications, we have no clue
about its contents. Further the application that created it, used its
own format, a format that is known to this application only. Today we
have the urge for analyzing and processing data found on the Grid (as
in typical OLAP applications), thus there is an undeniable need for
semantic description. Simply said, data without semantics is dead,
data with semantics lives. This statement leads naturally to the next
issue, persistency of data.

\subsubsection{Persistency of data}
Data stored without semantic information is lost (can not be reused),
because the semantics is originally only in the program code of the
application producing the data. Without the program the data is just
a sequence of bytes without meaning. With the usage of a framework
like xDGDL the data can be reused easily by any application
understanding the meaning of the data. A practical Java-based example
is given in \cite{parstorage}.

\subsubsection{Portability}
In a distributed environment parts of data can migrate from one
node/system/environment to another. On different hosting environments
naturally the data formats change. However when moving data from one
system to another, applications must still be able to read the data.
By the description of the format the data can be interpreted and can
be easily transformed to any proprietary format of the target machine
\cite{feitelson01}.

\subsubsection{Performance and efficiency} To enhance the bandwidth of
the IO media (to fight the famous IO bottleneck) it is the most
common technique to distributed the data among different nodes and/or
devices and perform the accesses in parallel. If the user has
knowledge about the available nodes or the application behavior she
can describe the distribution of the file to her needs. This can lead
to performance improvements especially if the user is aware of node's
performance, the given network latency, the network bandwidth to each
server, etc.

\subsection{The xDGDL specification
}
The Extensible Markup Language (XML) is the universal format for
structured documents and data on the Web. It describes a class of
data objects called XML documents and partially describes the
behavior of computer programs which process them.

XML documents are made up of storage units called entities, which
contain either parsed or unparsed data. Parsed data is made up of
characters, some of which form character data, and some of which form
markups. Markup encodes a description of the document's storage
layout and logical structure. XML provides a mechanism to impose
constraints on the storage layout and logical structure.

The structure of XML is fundamentally tree oriented. Therefore a
document can be modelled as an ordered, labelled tree, with a
document vertex serving as the {\it root} vertex and several {\it
child} vertices. Without the document vertex, an XML document may be
modelled as an ordered, labelled forest, containing only one root
element, but also containing the XML declaration, the doctype
declaration, and perhaps comments or processing instructions at the
root level.

To define the legal building blocks of an XML document, a DTD
(Document Type Definition) can be used. It defines the document
structure with a list of legal elements.

A DTD can be declared inline in your XML document, or as an external
reference.

It was a clear decision to choose XML as the basis for our framework
due to its undeniable success within the Internet community and its
acceptance as basis for beneath any standard movement in the Grid
community (e.g. WSDL \cite{wsdl}).

\subsection{The xDGDL document type definition}

In our framework a typical xDGDL description consists of the
following elements:
\begin{itemize}
    \item {\bf Document Root} The root of the document
    specifies the version and timestamp of the file of
    the XML description.
    \item {\bf Island} Defines a logical unit with several
    servers distributed worldwide. This element resembles the
    collaboration of our simple Grid architecture given above.
    \item {\bf Server} Servers are physical machines identified
    by their host name. These servers denote data nodes.
    \item {\bf Devices} Devices are the disks holding the data
    on the specific server.
    \item {\bf View} The View element allows a specific
    distribution within the device.
    \item {\bf Block} The Block element specifies the number
    of bytes to write to the specific disk.
\end{itemize}

The complete DTD of xDGDL can be found in the Appendix.

\subsubsection{Document root}
The root of the document is described by the element {\tt
PARSTORAGE}. It has the attribute {\tt VERSION} that contains the
version of the document and the attribute {\tt TIMESTAMP} that
identifies the external name together with the logical file. Both
attributes are mandatory.

The root element can contain several child elements. The {\tt
PROCESSORS} and the {\tt ALIGN} children are optional. The following
child elements are possible:
\begin{itemize}
    \item {\tt PROCESSORS} describes the named processor arrays. A document may contain zero or more processor array definition, which are normally derived from the HPF definition.
    \item {\tt TYPE} describes the data types and variables stored in
        the logical file. Types enhance the quality of stored data. They allow to define
        the meaning of the information stored. This leads to the fact that
        not only the program that stored the data can use them. Every program that
        understands the type information of the data can use the stored bytes.
        Because of these meta information it is also possible to migrate data
        from one machine to another.
        There must be at least one {\tt TYPE} element in the document.
    \item {\tt ALIGN} describes the alignments of the variables.
    \item {\tt ISLAND} describes the physical view of the file.
\end{itemize}
Example:
\begin{verbatim}
<PARSTORAGE VERSION="1.0"
            TIMESTAMP="testfile_twoserver">
    <TYPE>
        ...
    </TYPE>
    <ISLAND NAME="pri.univie.ac.at">
        ...
    </ISLAND>
</PARSTORAGE>
\end{verbatim}

\subsubsection{Island}
The {\tt ISLAND} describes several server interconnected together.
These servers can be distributed across the Grid. The island is
identified by an island name. The {\tt ISLAND} consists of one or
more servers. At least one server is needed to write the file
sequential to that server. The number of servers are received from
the number of child present. Example:
\begin{verbatim}
<ISLAND NAME="pri.univie.ac.at">
    <SERVER HOST="vipios.pri.univie.ac.at">
    </SERVER>
</ISLAND>
\end{verbatim}

\subsubsection{Server}
The {\tt SERVER} identifies uniquely a node. It has an attribute
called {\tt HOST} which mirrors the name of the server.

The {\tt SERVER} element consists of one or more {\tt DEVICE}
elements. At least one must be present for each server to know how
the file should be distributed on the several disks. For this purpose
the number of available devices on a specific server should be known.

Example:
\begin{verbatim}
<SERVER HOST="vipios.pri.univie.ac.at">
    <DEVICE DEVICE_ID="/dev/vda1">
    </DEVICE>
</SERVER>
\end{verbatim}

\subsubsection{Device}
Devices are the disks holding the data on the specific server. On one
{\tt SERVER} there could be more than one physical device. The server
can have a RAID system for example with several disks connected onto
it. The devices need not be physical, even a mounted NFS device on
another server could be a device which could be accessed from a
processing node. Although there can be many devices on a specific
server, in most cases there will be only one device available.

The {\tt DEVICE} element consists of the attribute {\tt DEVICE\_ID}
only, which specifies the physical device on the system. To describe
the structure of file parts to be written to disk, a {\tt VIEW} is
used. If there is no {\tt VIEW} defined we expect that the file
should be written sequential by the "first" logical server and the
"first" logical disk on this server.

Example:
\begin{verbatim}
<DEVICE DEVICE_ID="/dev/vda1">
    <VIEW SKIP_HEADER="0" SKIP="7">
    </VIEW>
</DEVICE>
\end{verbatim}

\subsubsection{View}
The {\tt VIEW} element is the link between logical, physical and
application view. It is responsible for transforming the internal
structure of the data layout to application programs.

A specific distribution is expressed by a {\tt VIEW} element. The
{\tt VIEW} needs to correspond to the servers available. The {\tt
NOVIEW} elements marks that there is no {\tt VIEW} element available.
If {\tt NOVIEW} is the only available child, the pointer to the
access-descriptor is set to NULL and therefore the file will be
written sequentially onto the disk. At least a {\tt VIEW} or a {\tt
NOVIEW} element has to be present.

The {\tt VIEW} consists of the {\tt SKIP\_HEADER} attribute that
describes how many header bytes are skipped at the beginning of the
data block and the {\tt SKIP} attribute that defines the number of
bytes to be skipped viewer units.

The {\tt VIEW} element consists of one or more {\tt BLOCK} elements.
Theoretically there can be an infinite number of {\tt BLOCK}
elements, but at least one is needed. The {\tt BLOCK} itself can have
another {\tt VIEW} element within itself.

Example:
\begin{verbatim}
<VIEW SKIP_HEADER="0" SKIP="7">
    <BLOCK OFFSET="0" REPEAT="3" COUNT="5" STRIDE="7">
        <BYTEBLOCK/>
    </BLOCK>
</VIEW>
\end{verbatim}

\subsubsection{Block}
The {\tt BLOCK} element can have two types of childs. It can have a
{\tt BYTEBLOCK} element, which means, that either there are no more
{\tt VIEW} elements or it can consist of {\tt VIEW} elements which
have one or more {\tt BLOCK} elements themselves. This leads to a
recursive structure which allows arbitrary distribution. At least one
has to be present.

The {\tt BLOCK} element consists of the following attributes:
\begin{itemize}
    \item {\bf OFFSET} describes how many bytes should be skipped from the starting point of the current {\tt BLOCK}.
    \item {\bf REPEAT} describes how often the {\tt BLOCK} should be read/written.
    \item {\bf COUNT} number of bytes to read/write at each {\tt BLOCK} operation.
    \item {\bf STRIDE} describes the number of bytes to skip at
    each {\tt BLOCK} operation.
\end{itemize}
Example of a regular distributed file onto 2 servers. The definition
on server 1
\begin{verbatim}
<BLOCK OFFSET="0" REPEAT="3" COUNT="5" STRIDE="7">
    <BYTEBLOCK/>
</BLOCK>
\end{verbatim}
corresponds to the definition on server 2:
\begin{verbatim}
<BLOCK OFFSET="5" REPEAT="3" COUNT="7" STRIDE="5">
    <BYTEBLOCK/>
</BLOCK>
\end{verbatim}

\begin{figure}
  \begin{center}
    \includegraphics[scale = .55]{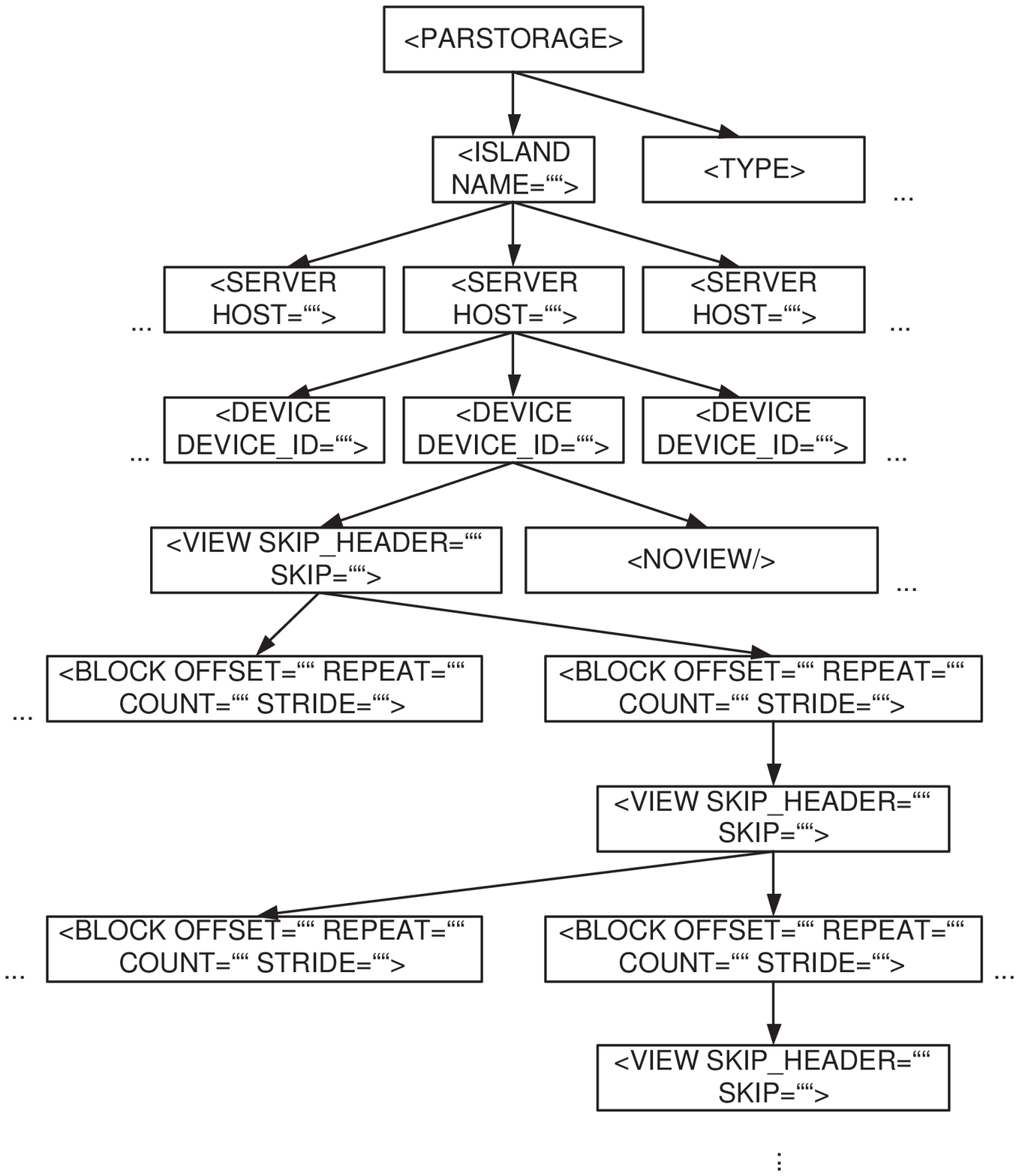}
   \caption{Example of a xDGDL tree}
  \end{center}
\end{figure}

\subsection{xDGDL examples}
The following three examples show several possibilities that the
xDGDL description provides. To depict the mapping between the
internal structure and the xDGDL description two figures are attached
to each example. The first figure shows a graphical tree
representation of the underlying XML structure and the second figure
the data distributed onto different servers.

\subsubsection{A regularly distributed, two-server example}
The first example introduces the structure of the xDGDL description.
It uses two servers and writes data in round robin fashion to the
local disks on each server: vipios.pri.univie.ac.at and
vipclus9.pri.univie.ac.at.

It is also possible to use more than one block. We would call this
an interleaved distribution. The interleaved distribution divides the
file into two parts. The first part is distributed on block one on
server one and block one on server two. The second part is distributed
on block two on server one and block two on server two.

The finer the granularity of the distribution gets, the more complex
the structure grows.\footnote{Beside this it is not wise to use a
fine granularity for small files as the overhead of parsing the
descriptor gets to large. In case of small files it would also lead
to the situation that the description file is probably bigger than
the files to write.}

We suppose that server one writes more data to the disk. The factor
is 5:7. (Please note it is an artificial example of minor practical
relevance!)

The xDGDL representation of the regular, two-server example:
\begin{verbatim}
<?xml version="1.0" encoding="ISO-8859-1"?>
<!DOCTYPE PARSTORAGE SYSTEM "XDGDL.dtd">
<PARSTORAGE VERSION="1.0"
            TIMESTAMP="testfile_regular">
 <TYPE>
  <ETYPE TYPE="CHAR" LENGTH="1"/>
 </TYPE>
 <ISLAND NAME="island1.pri.univie.ac.at">
  <SERVER HOST="vipios.pri.univie.ac.at">
   <DEVICE DEVICE_ID="/dev/vda1">
    <VIEW SKIP_HEADER="0" SKIP="7">
     <BLOCK OFFSET="0" REPEAT="3"
            COUNT="5" STRIDE="7">
      <BYTEBLOCK/>
     </BLOCK>
    </VIEW>
   </DEVICE>
  </SERVER>
  <SERVER HOST="vipclus9.pri.univie.ac.at">
   <DEVICE DEVICE_ID="/dev/vda1">
    <VIEW SKIP_HEADER="0" SKIP="0">
     <BLOCK OFFSET="5" REPEAT="3"
            COUNT="7" STRIDE="5">
      <BYTEBLOCK/>
     </BLOCK>
    </VIEW>
   </DEVICE>
   </SERVER>
 </ISLAND>
</PARSTORAGE>
\end{verbatim}

\begin{figure*}
  \begin{center}
    \includegraphics[scale = .8]{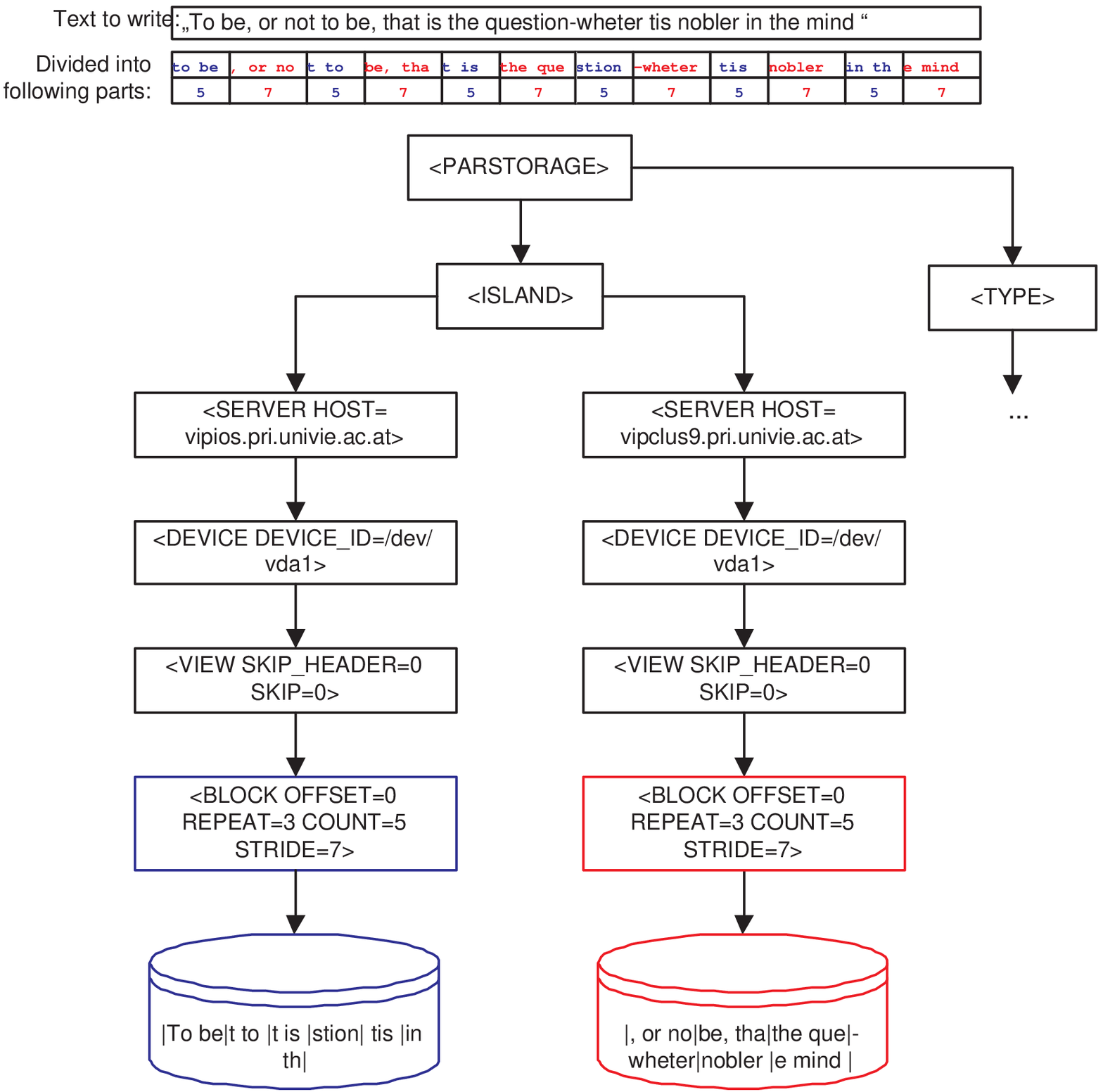}
   \caption{Tree representation of a regular distributed, two-server xDGDL distribution} \label{xParStorage_regular_tree}
  \end{center}
\end{figure*}

A graphical view of the regular distributed, two server example can
be seen in Figure \ref{xParStorage_regular_tree}


\subsubsection{A regular distributed, nested three-server example}
The last example handles three server. Beside the extension to three
servers it is also the one that shows a nested description. The
recursion depth itself is not limited.

The nested description gives the user an unrestricted flexibility to
express any data distribution.

The xDGDL description of a regular distributed, nested three-server
distribution:
\begin{verbatim}
<?xml version="1.0" encoding="ISO-8859-1"?>
<!DOCTYPE PARSTORAGE SYSTEM "XDGDL.dtd">
<PARSTORAGE VERSION="1.0"
            TIMESTAMP="regular_multilevel">
 <TYPE>
  <ETYPE TYPE="CHAR" LENGTH="1"/>
 </TYPE>
 <ISLAND NAME="island3.pri.univie.ac.at">
  <SERVER HOST="vipios.pri.univie.ac.at">
   <DEVICE DEVICE_ID="/dev/vda1">
    <VIEW SKIP_HEADER="0" SKIP="12">
     <BLOCK OFFSET="0" REPEAT="2"
            COUNT="1" STRIDE="12">
      <VIEW SKIP_HEADER="0" SKIP="0">
       <BLOCK OFFSET="0" REPEAT="3"
              COUNT="5" STRIDE="7">
        <BYTEBLOCK/>
       </BLOCK>
      </VIEW>
     </BLOCK>
    </VIEW>
   </DEVICE>
  </SERVER>
  <SERVER HOST="vipclus9.pri.univie.ac.at">
   <DEVICE DEVICE_ID="/dev/vda1">
    <VIEW SKIP_HEADER="0" SKIP="12">
     <BLOCK OFFSET="0" REPEAT="2"
            COUNT="1" STRIDE="12">
      <VIEW SKIP_HEADER="0" SKIP="0">
       <BLOCK OFFSET="5" REPEAT="2"
              COUNT="7" STRIDE="12">
        <BYTEBLOCK/>
       </BLOCK>
      </VIEW>
     </BLOCK>
    </VIEW>
   </DEVICE>
  </SERVER>
  <SERVER HOST="vipclus10.pri.univie.ac.at">
   <DEVICE DEVICE_ID="/dev/vda1">
    <VIEW SKIP_HEADER="0" SKIP="0">
     <BLOCK OFFSET="29" REPEAT="2"
            COUNT="12" STRIDE="29">
      <BYTEBLOCK/>
     </BLOCK>
    </VIEW>
   </DEVICE>
  </SERVER>
 </ISLAND>
</PARSTORAGE>
\end{verbatim}

\begin{figure*}
  \begin{center}
    \includegraphics[scale = .8]{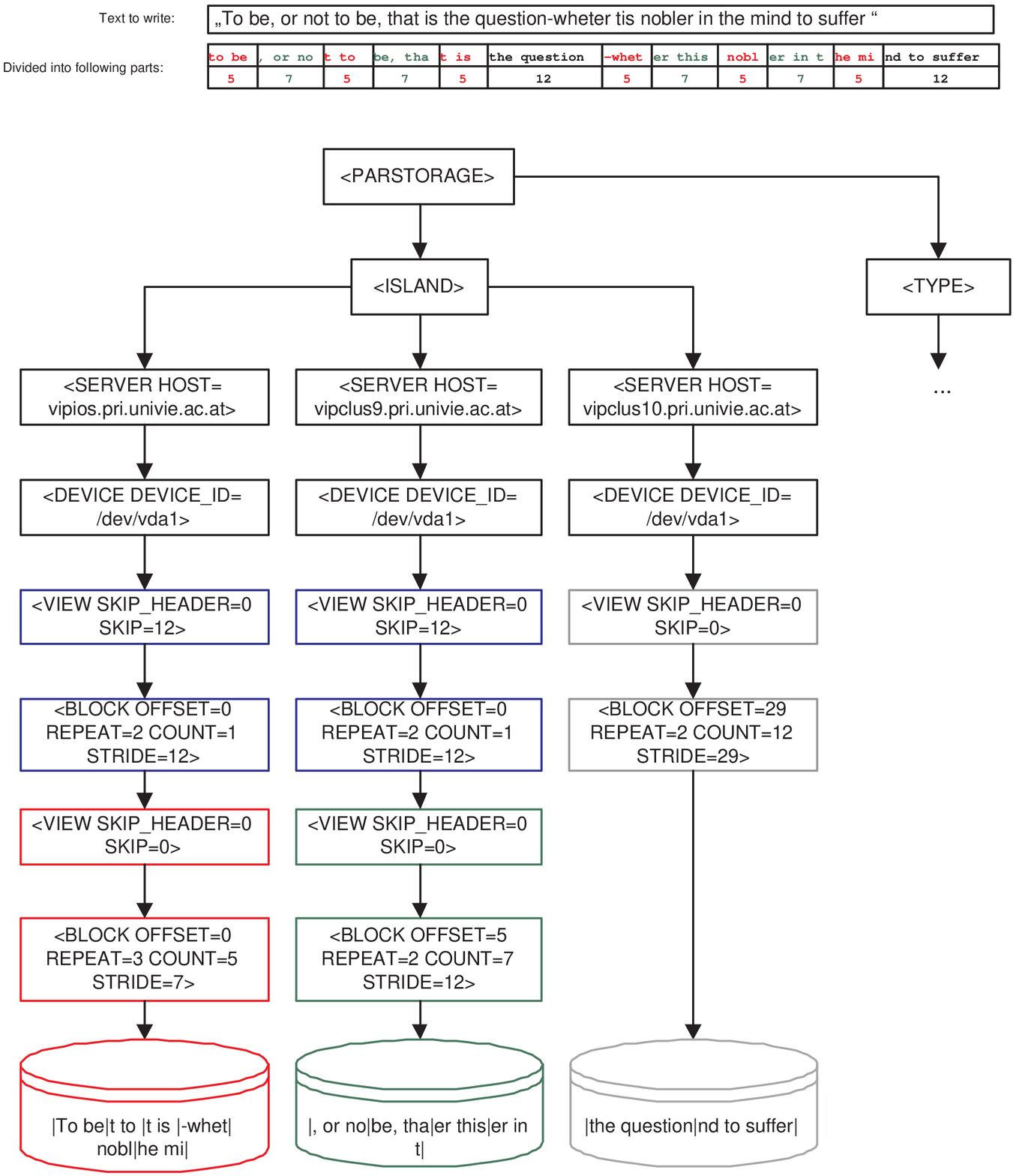}
   \caption{Tree representation of a regular distributed, nested three-server xDGDL distribution} \label{xParStorage_regular_multilevel_tree}
  \end{center}
\end{figure*}

A graphical view of the regular distributed, nested three-server
example can be seen in Figure
\ref{xParStorage_regular_multilevel_tree}

\section{An Application of xDGDL}

\subsection{The ViPIOS island}

ViPIOS - the Vienna Parallel Input Output System - is an I/O system
that tries to solve the well-known I/O bottleneck of high-performance
computing \cite{Brezany1996,es-vipios-europar98}. ViPIOS was originally designed
as a client-server system satisfying parallel I/O needs of high
performance applications. Due to the requirements of the Datagrid
initiative ViPIOS was extended to Meta-ViPIOS, which harnesses
distributed I/O resources \cite{fuerle00}.

A {\em ViPIOS island} (resembling roughly a collaboration within our
Grid architecture) can be seen as a logically independent system,
residing on a defined set of processing nodes. Conventionally this is
a typical cluster system, but it can also be an arbitrary set of
world-wide distributed machines. An island comprises an arbitrary
number of ViPIOS servers processing the I/O requests of connected
applications. To reach such an island the client needs to know the
hostname (or IP-address) of a dedicated connection server responsible
for that island (for more information see \cite{islands02}).

An island provides several interfaces; beside the native interface,
an MPI-IO interface (ViMPIOS), a HPF/VFC (Vienna Fortran Compiler)
interface as well as a Unix file access interface (ViPFS) are
supported.

The system defines two modes to describe the distribution of a file.
By default the automatic modes allows ViPIOS to decide how to
distribute the given file among the available servers. The user
guided modus in contrast let the user decide how to distribute the
file. In this modus a xDGDL file describes the distribution of a
given file.

\begin{figure}
  \begin{center}
    \includegraphics[scale = .45]{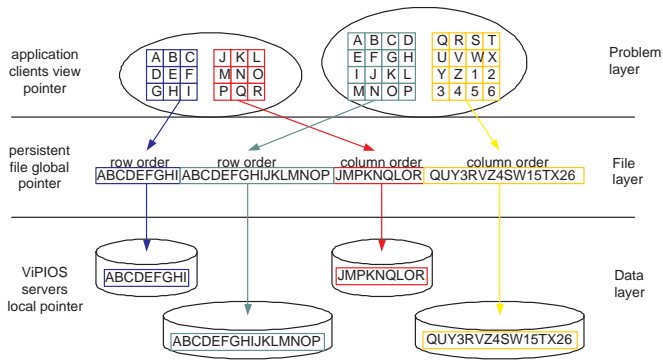}
   \caption{\label{chViPIOS_layers} Different point of views: The ViPIOS layers}
  \end{center}
\end{figure}

ViPIOS provides a data independent view of the stored data to the
application process. It is based on a three-tier model. The three
specific ViPIOS layers are the following (see Figure
\ref{chViPIOS_layers}):

\begin{itemize}
    \item {\bf Problem layer.} Defines the problem specific data
    distribution among the cooperating parallel processes (View
    file pointer).
    \item {\bf File layer.} Provides a composed view of the
    persistently stored data in the system (Global file pointer).
    \item {\bf Data layer.} Defines the physical data distribution
    among the available disks (Local file pointer).
\end{itemize}

The three tier architecture allows ViPIOS to be completely logical
data independent between the problem and the file layer as well as to
be physical independent between the file and data layer.

\subsection{The ViPIOS interfaces}
ViPIOS provides a range of interfaces to support a wide variety of
applications. The interfaces are supported by interface modules to
allow flexibility and extendibility. Up to now we implemented the
following modules:
\begin{itemize}
    \item HPF/VFC - High Performance Fortran interface based on the
    Vienna Fortran compiler
    \item ViMPIOS - a MPI-IO interface
    \item ViPFS - ViPIOS distributed file system
    \item ViPIOS proprietary interface for some specialized modules
\end{itemize}

In the context of this paper we concentrate on the novel ViPFS, that
allows both the casual and the experienced user to use ViPIOS in form
of a distributed file system.

\subsection{ViPFS}
Basically ViPFS is a library which overloads the standard file calls
in UNIX. This methods allows users easily and efficiently to employ
transparently services provided by ViPIOS. Thus all Unix tools for
file accesses can be used without recompiling. The idea is to
redirect the calls with "conventional" data files to the standard I/O
library and to redirect the calls with ViPFS data files to the ViPIOS
system. This approach is similar to PVFS \cite{ligon:pfs}.

Beside the overloaded Unix interface ViPFS also provides a
C-Interface, which can be linked with C-programs. This interface
provides nearly the same functionality as the standard I/O interface.

For users it is very easy to define the meta information for the data
file in focus. A respective xDGDL file has to be created and stored
in the same directory as the data file, which has the same name as
the data file, but with the prefix "{\tt.vd.}"\footnote{The prefix
stands for \emph{ViPIOS description}}. With an open statement the
ViPFS library checks if there is a corresponding xDGDL file for the
given file. The prefixed dot is used because these files are not
visible with the common {\tt ls} command. It is also quite common to
use the dot for configuration files and to a certain extent the
"{\tt.vd.*}" files can be seen as configuration files. When it is
parsed, its is checked against the given data type definition (DTD).
If the file is erroneous or does not exist the respective data file
will be distributed with the standard distribution of ViPFS which is
a cyclic distribution among the available ViPIOS servers.

\subsubsection*{Copy Example}

The copy command is a simple example to show the transparent usage of
the ViPFS file system. In this example it is the intent to copy a
data file from a convention Unix file system to ViPFS and back.

The preconditions for using ViPFS are the following:
\begin{itemize}
    \item Start of ViPIOS

    \item Configuration of the ViPIOS configuration file ({\tt ViPIOS.conf}) that was set up
    in the environment. In our example we used: \\
\begin{verbatim}
MAX_APP 5 MAX_SRV_FILE 32 DATA_BUFLEN 4096
SRV_GROUP_NAME "vipios_server" SRVR_DEVICE_LIST 3
  /home/felder/ViPIOS/dev1/
  /home/felder/ViPIOS/dev2/
  /home/felder/ViPIOS/dev3/
VIP_DIR "/home/felder/vipios"

\end{verbatim}
    \item Setting of Unix environment variable that points to the ViPIOS configuration file \\
    (e.g. {\tt VIP\_CONF=/home/felder/vipios/ViPIOS.conf}). The environment could be set up
    with the command {\tt export}.
    \item Setting up the {\tt LD\_PRELOAD} environment variable. The variable must point
    to the {\tt vipfsinvoke.so} shared object. In our example we set it up as follows:
    {\tt export LD\_PRELOAD=/home/felder/vipfs/vipfsinvoke.so}
\end{itemize}

After these steps the ViPFS can be used similar to an NFS mounted
device. The user uses standard Unix calls only for writing and
reading files. Internally all I/O calls on the specified directory
(VIP\_DIR) are passed to the ViPFS library. Therefore all the Unix
commands that use the standard I/O calls can be used with ViPFS.

In case of the example above the user can copy a data file simply by
the commands shown in figure \ref{copyfile}

\begin{figure*}
  \begin{center}
\begin{verbatim}
    felder@vipios:~/vipfstests > ls -al .vd.*
    -rw-r-----   1 felder   users  1177 Oct 14  2001 .vd.testfile

    felder@vipios:~/vipios > cp testfile /home/felder/vipios   # copy in
    felder@vipios:~/vipios > cp /home/felder/vipios/testfile . # copy out

    felder@vipios:~/vipios > ls -l /home/felder/vipios
    total 0
    -rw-r--r--   1 felder   users   0   Oct 14  2001 testfile
\end{verbatim}
\end{center}\caption{ViPFS copy of a data file}
\end{figure*}\label{copyfile}

As we did not overload the {\tt ls} command the user can only see a
file with 0 bytes within the VIP\_DIR. This is due to the fact that
the file is not really copied into the directory. For transparency to
the user ViPFS generates a 0-byte file to provide the user with the
information which files are currently distributed on the system.

In the first line we print out all .vd.* files. In our example
only one distribution file is present. We used the distribution
file presented in \ref{xParStorage_regular_multilevel_tree}. That
means, that the testfile was distributed among three servers with
one device on each server. If we did not declare a .vd. file the
testfile would have been written sequentially to the first disk on
the current server.

\section{Conclusions and Future Work}\label{conclusions}

We presented xDGDL, an XML language for storing meta information
for distributed files on the Grid. The proposed XML approach acts
in the system in two ways; on one hand it provides a user
interface to specify the contents (semantical information) and the
layout (physical information) of the file, on the other hand it is
the expressive mechanism within the system to administer the
distribution information of the files stored in the file system
across several sites on the Grid. We showed a practical
prove-of-concept implementation by the ViPFS distributed file
system.

The xDGDL language is the starting point for a new way of defining
data access paths on the Grid. We work on a research project to
define Grid I/O patterns, which allow to define I/O data streams on
the Grid easily. A stream can be seen as a graph where the vertices
are modules, which are instantiated from Grid I/O patterns, and the
edges are the data streams. Data is moved along such streams and
carries along from vertex to vertex its self-describing information
based on the xDGDL language. This allows the modules, which in fact
are active I/O resources (Grid fabrics), as distributed file systems,
database systems, etc., to interpret and to process the data. We work
on a method for the automatic generation of such Grid I/O graphs
based on heuristic methods \cite{blackboard}.

\subsection*{Acknowledgement}
The work described in this paper was partly supported by the Special
Research Program SFB F011 AURORA of the Austrian Science Fund.

\section*{Appendix: xDGDL DTD}\label{xParStorageDTD}
\begin{verbatim}
<?xml version="1.0" encoding="ISO-8859-1"?>

<!-- (c) Andras Belokosztolszki-->
<!-- 2000 -->
<!-- (c) Rene Felder -->
<!-- 2001 -->


<!ELEMENT PARSTORAGE
  (PROCESSORS*,TYPE+,ALIGN*,ISLAND)>
<!ATTLIST PARSTORAGE VERSION CDATA #REQUIRED>
<!ATTLIST PARSTORAGE TIMESTAMP ID #REQUIRED>

<!-- processors -->
<!ELEMENT PROCESSORS (PROC_DIMENSION)+>
<!ATTLIST PROCESSORS NAME CDATA #REQUIRED>
<!ELEMENT PROC_DIMENSION EMPTY>
<!ATTLIST PROC_DIMENSION LOWER CDATA "1">
<!ATTLIST PROC_DIMENSION UPPER CDATA #REQUIRED>


<!-- hpf data structure -->
<!-- Intrinsic Data Types -->
<!ELEMENT TYPE (ETYPE|ARRAY|TYPE)+>
<!ATTLIST TYPE TYPENAME CDATA #IMPLIED>
<!ATTLIST TYPE NAME CDATA #IMPLIED>

<!ELEMENT ETYPE EMPTY>
<!ATTLIST ETYPE TYPE CDATA #REQUIRED>
<!ATTLIST ETYPE LENGTH CDATA #REQUIRED>
<!ATTLIST ETYPE NAME CDATA #IMPLIED>

<!-- Arrays -->
<!ELEMENT ARRAY (TYPE, DIMENSION+)>
<!ATTLIST ARRAY NAME CDATA #IMPLIED>
<!ATTLIST ARRAY MAJOR (ROW|COLUMN) "ROW">
<!ATTLIST ARRAY DISTRIBUTE_ONTO CDATA #IMPLIED>

<!ELEMENT DIMENSION EMPTY>
<!ATTLIST DIMENSION LOWER CDATA "1">
<!ATTLIST DIMENSION UPPER CDATA #REQUIRED>
<!ATTLIST DIMENSION DISTRIBUTE
  (BLOCK|CYCLIC|NO) #IMPLIED>
<!ATTLIST DIMENSION DIST_SKALAR CDATA "1">


<!-- Alignment -->
<!ELEMENT ALIGN EMPTY>
<!ATTLIST ALIGN WHAT CDATA #REQUIRED>
<!ATTLIST ALIGN WITH CDATA #REQUIRED>


<!-- data distribution in this file -->
<!-- Model Island-Descriptor -->
<!ELEMENT ISLAND (SERVER*)>
<!ATTLIST ISLAND NAME CDATA #REQUIRED>

<!-- Model Server-Descriptor -->
<!ELEMENT SERVER (DEVICE*)>
<!ATTLIST SERVER HOST CDATA #REQUIRED>

<!-- Model Device-Descriptor -->
<!ELEMENT DEVICE (VIEW|NOVIEW)>
<!ATTLIST DEVICE DEVICE_ID CDATA #REQUIRED>

<!-- Model Access-Descriptor -->
<!ELEMENT VIEW (BLOCK+)>
<!ATTLIST VIEW SKIP_HEADER CDATA #REQUIRED>
<!ATTLIST VIEW SKIP CDATA #REQUIRED>

<!ELEMENT BLOCK (VIEW|BYTEBLOCK)>
<!ATTLIST BLOCK OFFSET CDATA #REQUIRED>
<!ATTLIST BLOCK REPEAT CDATA #REQUIRED>
<!ATTLIST BLOCK COUNT CDATA #REQUIRED>
<!ATTLIST BLOCK STRIDE CDATA #REQUIRED>
<!ELEMENT BYTEBLOCK EMPTY>

\end{verbatim} 

\bibliographystyle{plain}
\bibliography{schiki}

\end{document}